\documentclass[twocolumn,secnumarabic,amssymb, nobibnotes,aps,prl]{revtex4-1}
\usepackage{graphicx}
\usepackage{textcomp}
\usepackage{amsmath}
\usepackage{commath}
\usepackage{color}

\newcommand{\sg}[1]{}
\renewcommand{\sg}[1]{{\color{black}{#1}}} 

\begin{document}
	\title{Symmetry of the phonon landscape of twisted kagomes lattices \\ across the duality boundary}
	\author{Stefano Gonella}
	\email{sgonella@umn.edu}
	\affiliation{Department of Civil, Environmental, and Geo- Engineering\\ University of Minnesota, Minneapolis, MN 55455, USA\\}
	
	\begin{abstract}
		In this work, we investigate the symmetry of the phonon landscape of twisted kagome lattices across their duality boundary. The study is inspired by recent work by Fruchart et al. [Nature, 577, 2020], who specialized the notion of duality to the mechanistic problem of kagome lattices and linked it to the existence of duality transformations between configurations that are symmetrically located across a critical point in configuration space. Our first goal is to elucidate how the existence of matching phonon spectra between dual configurations manifest in terms of observable wavefield characteristics. To this end, we explore the possibility of aggregating dual kagome configurations into bi-domain lattices that are geometrically heterogeneous but retain a dynamically homogeneous response. Our second objective is to extend the analysis to structural lattices of beams, which are representative of realistic cellular metamaterials. We show that, in this case, the symmetry of the phonon landscape across the duality boundary is broken, implying that the conditions for dual behavior do not merely depend on the geometry, but also on the dominant mechanisms of the cell, suggesting a dichotomy between geometric and functional duality. 
		\vspace{0.4cm}
	\end{abstract}
	
	\maketitle
	
	Kagome lattices are a special family of two-dimensional periodic structures whose unit cell consists of two triangles (solid or hollow) connected at one vertex. 
	The most prominent example is the regular kagome lattice, in which two equilateral triangles are relatively rotated by $180^o$. 
	Kagome lattices have received growing attention in the metamaterials literature, with a variety of studies addressing their mechanical properties~\cite{hyun_torquato_2002, Fleck_Lattices_JMPS_2007, Simons-Fleck_Imperfections_JAM_2008, Zhang_Mechanical_Lattices, Arabnejad-Pasini_Homogenization_IJMS_2013} and wave propagation characteristics~\cite{Phani_2006,Schaeffer-Ruzzene_Magneto-kagome_JAP_2015,Riva_Tunable_kagome_2018,Chen-et-al_QSHE-kagome_PRB_2018}.
	They have also been studied as viable mechanical models for biological materials, such as cartilage~\cite{Cohen_Kagome-cartilage_J-BioPhys_2014}, and as effective configurations for medical implants~\cite{Pasini_Femoral-implant_JMBBM_2013}. 
	Twisted kagome lattices represent a variation on the regular kagome paradigm, in which the triangles are arbitrarily rotated and the twist introduces dramatic changes in the effective properties ~\cite{sun2012surface}, including a switch between bulk and shear as the dominant stiffness-providing mechanism.  
	
	Many peculiar properties of kagome lattices stem from the fact that they belong to the family of Maxwell lattices~\cite{Maxwell1864,Calladine1978}, which contain an equal number of degrees of freedom and constraints in the bulk and are on the verge of mechanical instability~\cite{lubensky2015phonons,Souslov2009,Mao2011a}. 
	Recently, a spur of interest has surrounded phenomena rooted in the topology of these structures. Topological kagome lattices, obtained through a deformation of the regular kagome geometry, have been shown to display a topological polarization~\cite{kane2014topological}, which can be tuned by controlling the cell's twist and stretch~\cite{rocklin2017transformable} and results in floppy boundaries that support localized edge modes. In dynamics, it has been shown that the floppy boundaries  can support edge-confined phonons at finite frequencies~\cite{Ma-et-al_Topological-Waves_PRL_2018,Stenull-Lubensky_Topological-Phonons_PRL_2019,Zhou_Topo-Diode_PRB_2020} and topological polarization leads to asymmetric wave transport. 
	
	Recently, Fruchart et al. revisited the mechanics of twisted kagome lattices putting forth the notion of duality~\cite{Fruchart-Vitelli_Dualities_Nature_2020}, which can be interpreted as a behavioral symmetry in the parameter space of the twist angle, whereby configurations that are equidistant from a critical angle feature identical phonon spectra. In \cite{Fruchart-Vitelli_Dualities_Nature_2020}, this property is linked to the existence of a precise duality transformation between configurations on either side of the critical point. This study proposes a reflection on the implications of this powerful form of duality for the dynamics of cellular metamaterials. We first consider lattices of rods and we elucidate a series of geometric and spectral conditions under which duality can be established. We then explore the transportability of the concept to structural lattices of beams. This case mimics the realistic scenario of lattice materials obtained via fabrication techniques such as cutting or additive manufacturing, with implications of utmost relevance for engineering problems. 
	
	\begin{figure} [!htb]
		\centering
		\includegraphics[scale=0.5]{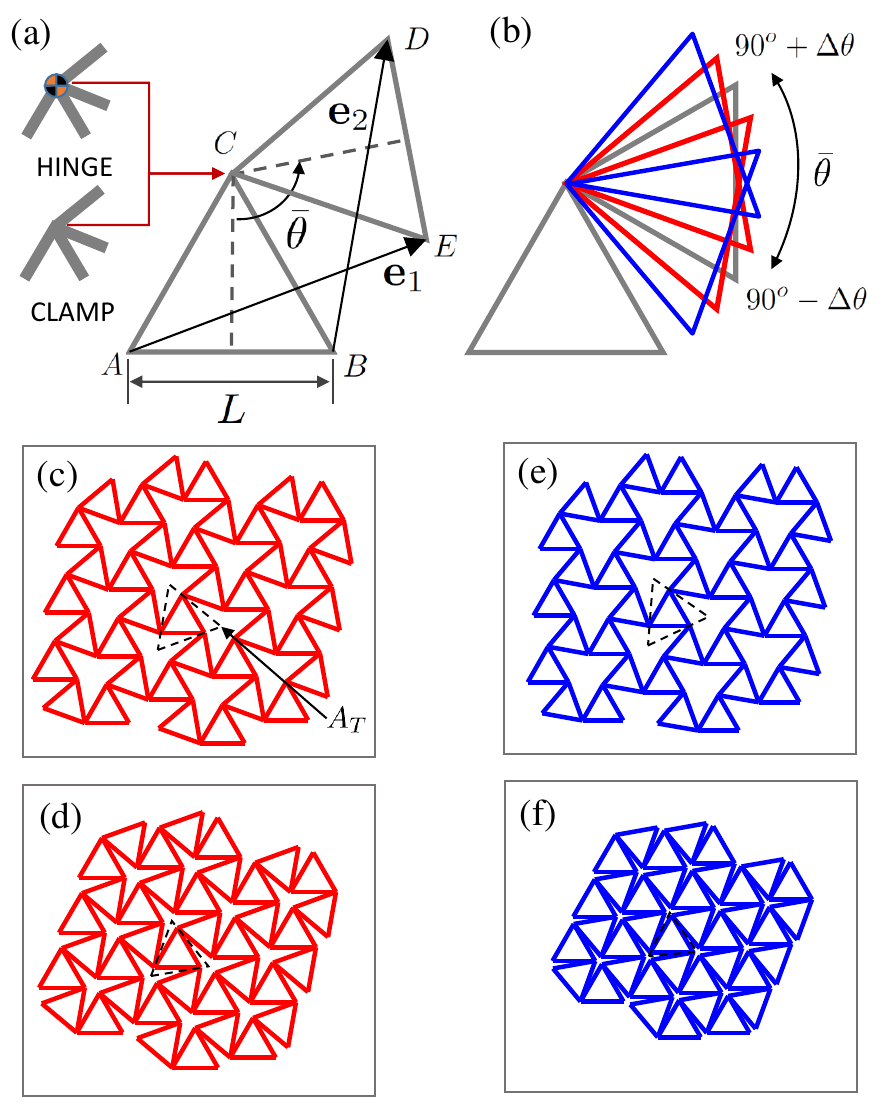}
		\caption{(a) Geometry of the twisted kagome lattice unit cell, showing primitive vectors and choice of hinge or clamp conditions. (b) Twisted kagome lattice family parameterized in terms of the twist angle $\bar{\theta}$. (c)-(d) Dual pair with $\bar{\theta}=105^o$ and $\bar{\theta}=75^o$. (e)-(f) Dual pair with $\bar{\theta}=110^o$ and $\bar{\theta}=70^o$.}
		\label{Geometry}
	\end{figure}
	
	The unit cell of a twisted kagome lattice, shown in Fig.~\ref{Geometry}(a), consists of two arbitrarily rotated equilateral triangles of side length $L$. Here, the  triangles are taken to be either trusses of rods connected by hinges, or frames of beams joined by internal clamps. In the former case, the bonds support only tension/compression and the perfect hinges allow free rotation of the rods, thus approximating ideal lattice conditions (although, here, the mass is not lumped at the sites). In the latter, the bonds support bending deformation and the clamps offer infinite resistance against the beams rotation. 
	The geometry is conveniently parameterized in terms of the twist angle $\bar{\theta}$, defined as the angle by which the top triangle is rotated about  point $C$, such that $\bar{\theta}=0^o$ yields two overlapping triangles and $\bar{\theta}=180^o$ returns the regular kagome cell. The primitive vectors are expressed as $\mathbf{e}_1=[L/2+L\sin(\bar{\theta}-30^o)] \, \textbf{i}_1+[\sqrt{3}L/2-L\cos(\bar{\theta}-30^o)] \, \textbf{i}_2$ and $\mathbf{e}_2=[-L/2+L\cos(\bar{\theta}-60^o)] \, \textbf{i}_1+[\sqrt{3}L/2+L\sin(\bar{\theta}-60^o)] \, \textbf{i}_2$, with $\textbf{i}_1, \, \textbf{i}_2$ Cartesian unit vectors. Two configuration pairs $90^o \pm \Delta \theta$ (Fig.~\ref{Geometry}(b)), symmetrically located across the $90^o$ critical point, form a \textit{dual pair}. Two pairs are denoted in red and blue and the corresponding color-coded lattices are shown in Fig.~\ref{Geometry}.(c)-(d) and~(e)-(f). 
	
\begin{figure} [t]
		\centering
		\includegraphics[scale=0.5]{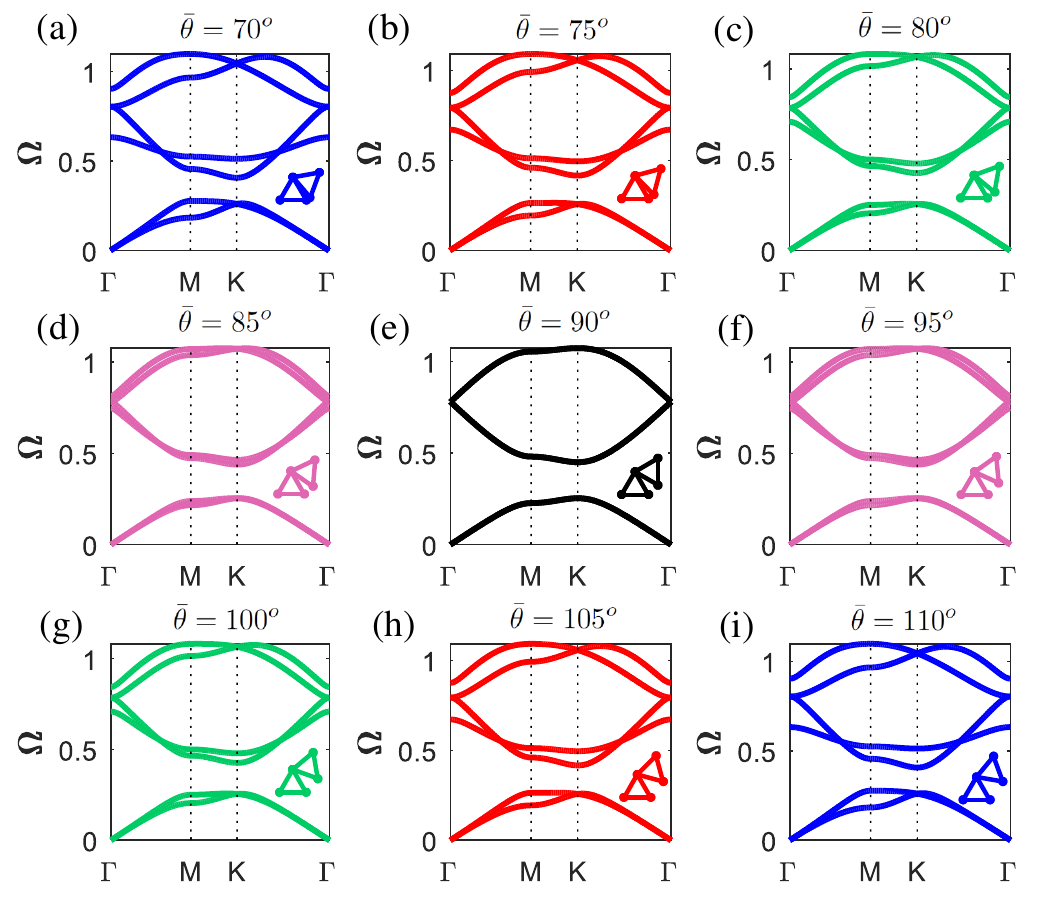}
		\caption{Band diagrams for lattices of rods connected by hinges for $\bar{\theta} \in [70^o \,\, 110^o]$, color-coded to highlight the dual pairs, for which \textit{matching} phonon spectra are obtained.}
		\label{BG_Rods}
	\end{figure}

	In Fig.~\ref{BG_Rods}, we plot the band diagrams for lattices of rods with $\bar{\theta} \in [70^o \,\, 110^o]$, color-coded to highlight the dual pairs. The frequency is normalized as $\Omega=\omega/\omega_{0R}$, where $\omega_{0R}=\pi/L\sqrt{E/\rho}$ is the first natural frequency of a rod of length $L$, Young's modulus $E$ and density $\rho$. We observe that dual cells feature identical phonon spectra, retrieving a signature of duality analogous to that presented in~\cite{Fruchart-Vitelli_Dualities_Nature_2020} (where, note, the lumped mass may lead to slightly different branches). 
	\begin{figure} [b]
		\centering
		\includegraphics[scale=0.45]{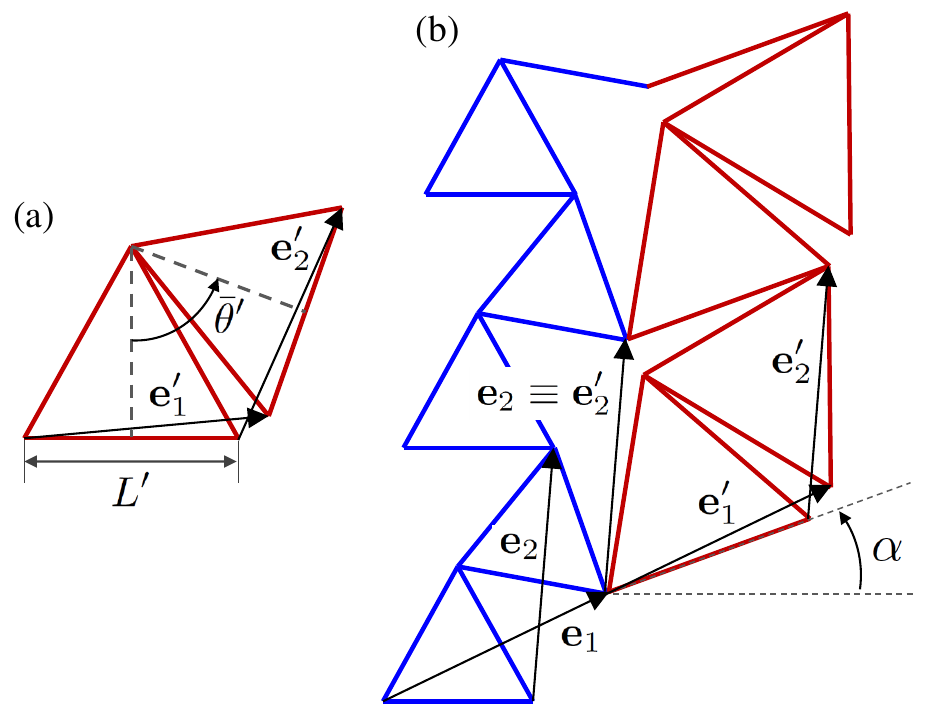}
		\caption{(a) Unit cell of lattice 2 with $\bar{\theta}'=70^o$, to be stitched to its $\bar{\theta}=110^o$ dual. (b) Stitching strategy showing compatibility constraints necessary to obtain a matching interface.} 
		\label{Two_lattices}
	\end{figure}
	Here, it is important to clarify how matching phonon landscapes between dual configurations translate into dimensional wavefield descriptors. In Fig.~\ref{BG_Rods}, the wavevector is sampled along the $\Gamma \textrm{M} \textrm{K} \Gamma$ path in (non-dimensional) reciprocal space. 
	Since the reciprocal base varies with $\bar{\theta}$, the dimensional wavevector $\mathbf{k}=k_x \mathbf{i}_1+k_y \mathbf{i}_2$ changes between configurations. Therefore, two corresponding points in the band diagrams of dual configurations denote plane waves with different wavelengths and propagation directions. Also, the phase velocity vector $\mathbf{c}^p=\omega \, \mathbf{k}/||\mathbf{k}||=c^p_x \mathbf{i}_1+c^p_y \mathbf{i}_2$  varies in magnitude and orientation and the directivity patterns are tilted between configurations (see SM). 
	
	An effective way to visualize the effects of duality within these constraints is by constructing a lattice composed of two subdomains meeting at an interface. 
	The assembly strategy is depicted in Fig.~\ref{Two_lattices}, assuming $\bar{\theta}=110^o$ and $\bar{\theta}'=70^o$ for lattices 1 and 2, respectively (quantities pertaining to lattice 2 are marked as $()' \, $). We recognize from Fig.~\ref{Two_lattices}(b) that, in order to have a matching interface, lattice 2 must feature a larger cell size $L'$ and rotate by an angle $\alpha$ and, along the interface, $\mathbf{e}_2$ and its counterpart $\mathbf{e}'_2$ must match (upon rotation by $\alpha$). Therefore, to find $L'$, it is sufficient to enforce $||\mathbf{e}_2|| =  ||\mathbf{e}'_2||$ and solve for $L'$; $\alpha$ is precisely the angle by which $\mathbf{e}'_2$ must rotate to overlap with $\mathbf{e}_2$, hence $\alpha=\cos^{-1}\left( \frac{\mathbf{e}_2 \cdot \mathbf{e}'_2}{||\mathbf{e}_2|| \,  ||\mathbf{e}'_2||} \right)$. 
	\begin{figure} [t]
		\centering
		\includegraphics[scale=0.53]{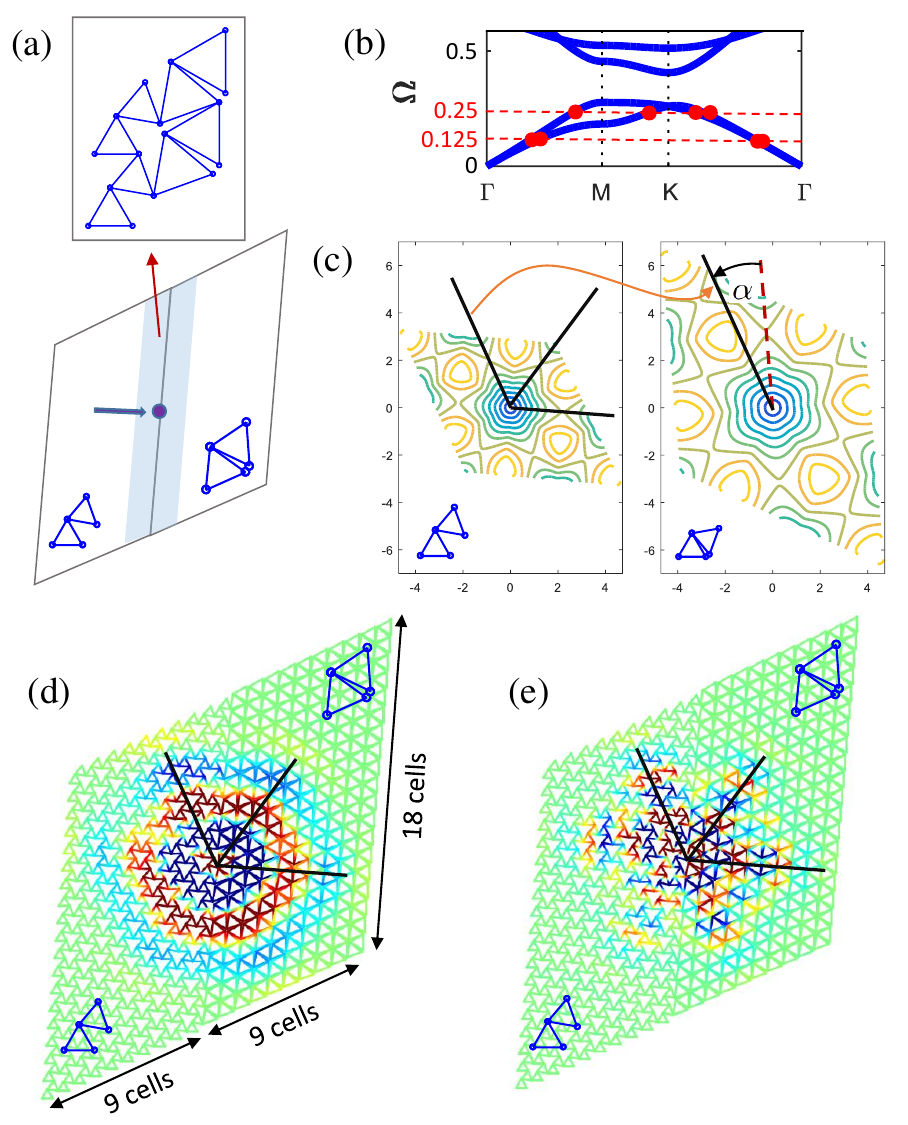}
		\caption{(a) Schematic of lattice of rods with $\bar{\theta}=110^o$ and $\bar{\theta}'=70^o$ subdomains, properly stitched with material compensation. (b) Detail of band diagram, with highlighted frequencies of excitation. (c) I mode iso-frequency contours for $\bar{\theta}=110^o$ and $\bar{\theta}=70^o$, showing inherent directivity mismatch between duals. (d) and (e) Snapshots of wavefields excited at $\Omega=0.125$ and $\Omega=0.25$, respectively (color proportional to horizontal displacement). The wavefields display \textit{matching} propagation characteristics in the two subdomains.}
		\label{Wavefields_Rods}
	\end{figure}
	Moreover, since the frequencies in the band diagrams are normalized by $\omega_{0R}$, which reflects the jump from $L$ to $L'$, the phonon correspondence is compromised across the interface. In order to compensate for this jump and preserve the phonons across the interface, we need to correct the material properties of lattice 2. Since $\omega_{0R} \propto 1/L\sqrt{E/\rho}$, this can be achieved by simply enforcing 
	$(E/\rho)'/(E/\rho)=(L'/L)^2$. 
	We consider now a domain comprising two dual subdomains, as shown in Fig.~\ref{Wavefields_Rods}(a). A 5-cycle tone burst excitation is applied at the center, perpendicular to the interface. In Fig.~\ref{Wavefields_Rods}(b) we show how the selected frequencies intersect the matching band diagrams of the dual configurations, predicting nearly-isotropic propagation at $\Omega=0.125$ and highly directional behavior at $\Omega=0.25$. 
	In Fig.~\ref{Wavefields_Rods}(c) we compare the iso-frequency contour lines of the lowest acoustic mode for the dual lattices. 
	We confirm that the inherent directivity landscapes are rotated by precisely $\alpha$. This directional mismatch is automatically compensated by the rotation operation performed while stitching the subdomains. Two snapshots of the resulting wavefields are shown in Fig.~\ref{Wavefields_Rods}(d) and (e). Remarkably, both wavefields display matching characteristics across the interface. 
	This result suggests that, by stitching dual kagome configurations with properly tailored material properties, it is possible to design lattices that are at once geometrically heterogeneous and dynamically homogeneous. 
	
	\begin{figure} [t] 
		\centering
		\includegraphics[scale=0.6]{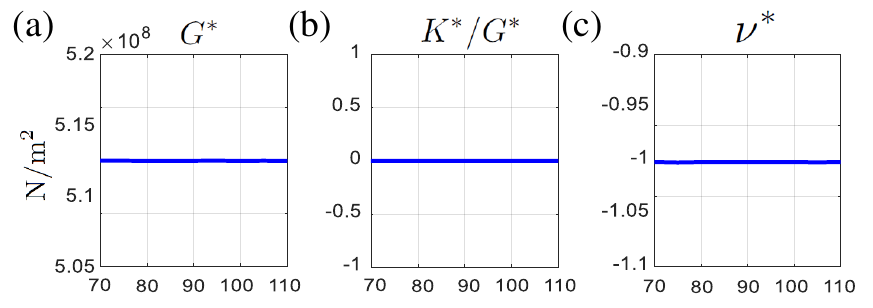}
		\caption{Effective moduli of lattices of rods vs. $\bar{\theta}$. (a) Effective shear modulus $G^*$, assuming $L= 1 \, \textrm{m}$, $\beta=1/15$, Young's modulus $E=70 \cdot 10^9 \, \textrm{N}/\textrm{m}^2$, Poisson's ratio $\nu=0.33$, density $\rho=2700 \, \textrm{Kg}/\textrm{m}^3$. (b) Ratio of effective bulk modulus $K^*$ and effective shear modulus $G^*$. (c) Effective Poisson's ratio $\nu^*$.}
		\label{Moduli_Rods} 
	\end{figure}

	From the band diagrams, we can estimate the effective elastic moduli. First, we determine the effective density $\rho^*=\rho \bar{\rho}$, where 
	$\bar{\rho}$ is the relative density, obtained dividing the area of a half cell occupied by solid by the total ``foot print" area $A_T$ (see Fig.~\ref{Geometry}(c)). Here, $\bar{\rho}=3Lh/A_T$, where $h=\beta L$ and $\beta$ is the rod's slenderness ratio. 
	We then determine numerically the phase velocities $c^p_S$ and $c^p_L$ of the acoustic modes in the long-wavelength limit ($||\textbf{k}|| \rightarrow 0$). By comparing $c^p_S$ and $c^p_L$ along $\Gamma \textrm{M}$ and $\Gamma \textrm{K}$, it is easy to verify that the long-wavelength behavior is isotropic for all $\bar{\theta}$. 
	Finally, the effective shear modulus $G^*$ and bulk modulus $K^*$ are computed from the relations for linear homogeneous and isotropic media: $c^p_S=\sqrt{G^*/\rho^*}$ and $c^p_L=\sqrt{(K^*+G^*)/\rho^*}$.  
	$G^*(\bar{\theta})$, plotted in Fig.~\ref{Moduli_Rods}(a) for a given material selection, is found to be 
	constant for all $\bar{\theta}$. As inferable from the plot of $K^*/G^*$ in Fig.~\ref{Moduli_Rods}(b), the procedure yields negligible bulk modulus values 
	(the methodology, which is sensitive to inaccuracies in the inference of $c^p_S$ and $c^p_L$, yields small enough values that can be effectively interpreted as $K^* \approx 0$). The effective Poisson's ratio $\nu^*$ is found from the relation $\nu^*=(K^*+G^*)/(K^*-G^*)$ for 2D plane-stress~\cite{Eischen-Torquato_Lattices_JAP_1993}. We see in Fig.~\ref{Moduli_Rods}(c) that $\nu^*$ approaches -1 (deep auxetic behavior) for all $\bar{\theta}$. These results match the theoretical properties of ideal twisted kagome lattices~\cite{sun2012surface}. 
	
\begin{figure} [t] 
		\centering
		\includegraphics[scale=0.6]{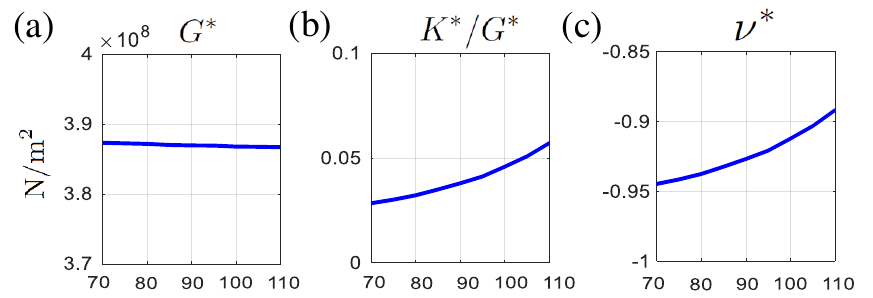}
		\caption{Effective moduli of lattices of beams vs. $\bar{\theta}$. (a) Effective shear modulus $G^*$. (b) Ratio between effective bulk modulus $K^*$ and $G^*$. (c) Effective Poisson's ratio $\nu^*$.}
		\label{Moduli_Beams_1over10} 
	\end{figure}

	We now switch our focus to lattices of beams, in which 
	the bonds are discretized with Timoshenko beam elements. 
	The Timoshenko model is robust against a wide spectrum of 
	$\beta=h/L$, including the limit of thick, shear-deformable beams (e.g., $\beta > 1/5$)~\cite{Phani_2006}. However, since our configurations feature extremely re-entrant angles, excessive $\beta$ values would be unrealistic, as the clamps would morph into platelets, challenging the validity of a beam model. Therefore, we will focus on slender beams ($\beta=1/10$ here). 
	\begin{figure} [t] 
		\centering
		\includegraphics[scale=0.5]{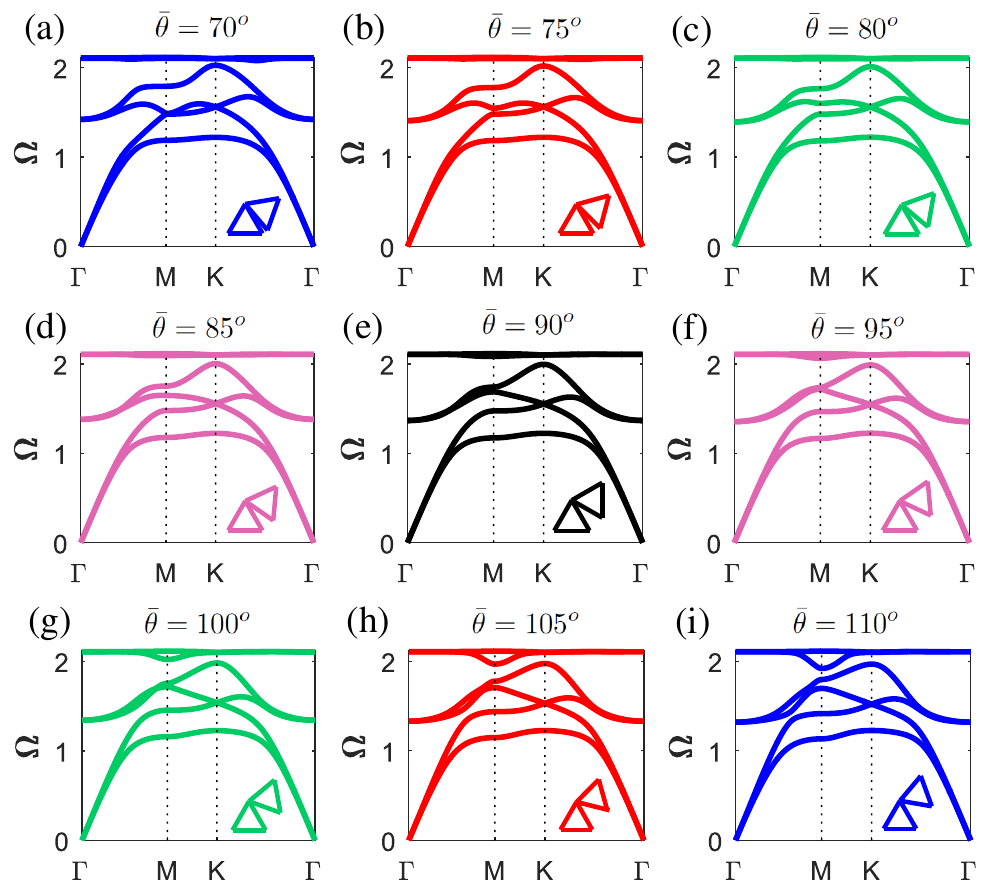}
		\caption{Band diagrams for lattices of beams for $\bar{\theta} \in [70^0 \,\, 110^o]$, color-coded to highlight the dual pairs. In this case, the phonon spectra of dual pairs are \textit{different}.}
		\label{BG_Beams_1over10} 
	\end{figure}
	The band diagrams for $\bar{\theta} \in [70^o \,\, 110^o]$ are shown in Fig.~\ref{BG_Beams_1over10}. 
The frequency is normalized by the first natural frequency of a simply-supported beam of length $L$: 
$\omega_{0B}=\pi^2/L^2\sqrt{E I/\rho A}$, where $A$ and $I$ are the cross-sectional area and second moment of area. Interestingly, the band diagrams of dual pairs are no longer identical, with differences that grow when we move away from the critical angle. In the SM, we show that these differences become more pronounced as $\beta$ increases and we depart from ideal lattice conditions. 
This result implies that the availability of matching phonon spectra does not merely depend on geometry but also on the specific mechanisms of deformation that the unit cell supports. While \textit{geometric duality} can be obtained by mere twisting, the emergence of \textit{functional duality} depends on other factors and is not automatically guaranteed.
The effective moduli are plotted in Fig.~\ref{Moduli_Beams_1over10}. 
We observe appreciable differences from the rods case. While $G^*$ is still almost constant, 
$K^*$ remains smaller than $G^*$ but is no longer negligible, and $\nu^*$ deviates from -1, albeit remaining in the deep auxetic regime. 
Moreover, the property landscape is no longer symmetric about the duality boundary. 

	\begin{figure} [b] %
		\centering
		\includegraphics[scale=0.65]{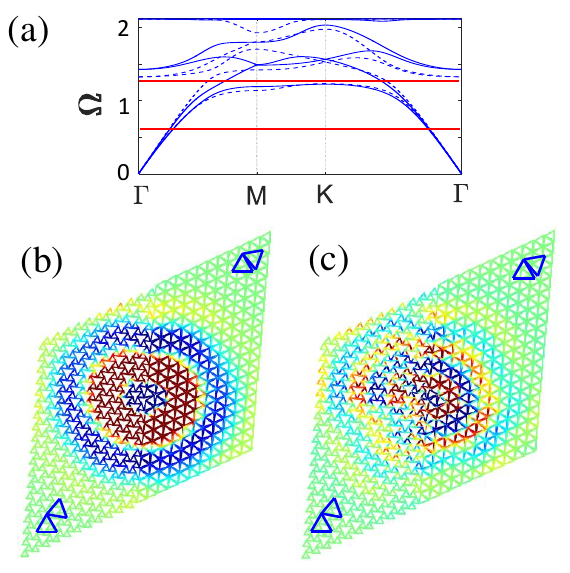}
		\caption{(a) Band diagrams for lattices of beams with $\bar{\theta}=110^o$ and $\bar{\theta}'=70^o$, superimposed with excitation frequencies highlighted. (b)-(c) Snapshots of wavefields excited by frequencies indicated in (a). The wavefields feature \textit{different} directionality and dispersive characteristics in the two subdomains.}
		\label{Wavefields_Beams_1over10} 
	\end{figure}

	We now revisit the lattice with 
	stitched subdomains as a frame of beams. 
	Since $\omega_{0B}  \propto 1/L\sqrt{E/\rho} \, \beta$ and we keep $\beta$ constant across the interface, it is again sufficient to select the 
	properties of lattice 2 such that $(E/\rho)'/(E/\rho)=(L'/L)^2$ to 
	establish identical spectral conditions across the interface. 
	The snapshots of the wavefields shown in Fig.s~\ref{Wavefields_Beams_1over10}(b)-(c), excited by the frequencies highlighted on the band diagrams of Fig.~\ref{Wavefields_Beams_1over10}(a), suggest that waves now propagate in the two subdomains with different directionality and levels of dispersion despite the nominal geometric duality of the configurations. 
	\begin{figure} [t]
	\centering
	\includegraphics[scale=0.7]{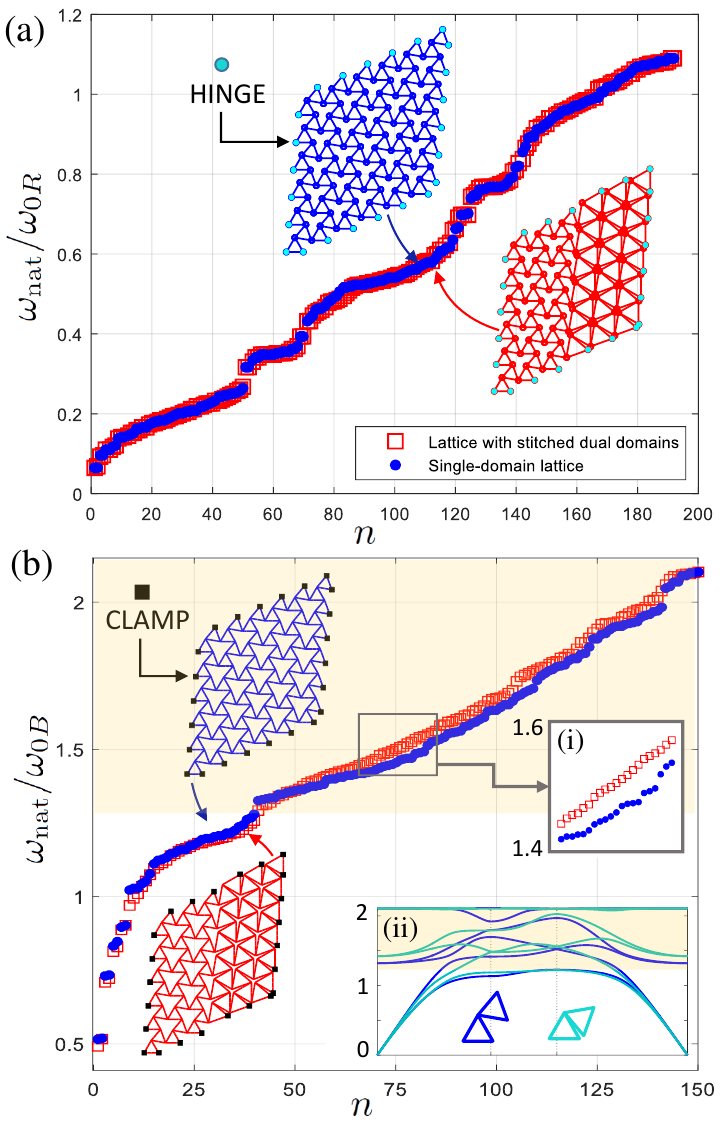}
	\caption{\sg{Comparison of natural frequencies between single-domain lattice and bi-domain lattice with stitched dual configurations. (a) Lattice of rods: the frequencies match across the set. (b) Lattice of beams: deviations are observed between the two scenarios; a detail of the region of deviation is given in inset (i). Inset (ii) recalls the difference in band diagram between the $\bar{\theta} = 110^o$ and $\bar{\theta} = 70^o$ configurations, color-coded in blue and turquoise, respectively. The yellow band in inset (ii) marks the interval where the branch morphologies depart from one another, corresponding to the region of the natural frequencies plot highlighted in yellow, where the frequency deviation is most pronounced.}}
	\label{Nat_freq_comparison}
	\end{figure}

	\sg{A powerful demonstration of the potential of geometrically heterogeneous but dynamically homogeneous bi-domain lattices can be obtained by looking at the vibration characteristics of finite structures. Additionally, this exercise further assesses the robustness of functional duality against relaxation of the ideality of the hinges. 
	In Fig.s~\ref{Nat_freq_comparison}(a) and (b), we compute the natural frequencies for a $6 \times 6$ (cell wise) lattice of rods with hinged perimeter nodes and for a corresponding lattice of beams with clamped boundaries, respectively. For each case, we compare two scenarios. In scenario 1 (color-coded in blue), $\bar{\theta} = 110^o$ everywhere. In scenario 2 (color-coded in red), we stitch two $3 \times 6$ dual strips with $\bar{\theta} = 110^o$ and  $\bar{\theta}' = 70^o$, respectively. 
	The corresponding natural frequencies 
	are plotted as blue circles and red squares (here, we plot the natural frequencies falling in the interval spanned by the band diagrams of Fig.s~\ref{BG_Rods} and~\ref{BG_Beams_1over10}, respectively). From Fig.~\ref{Nat_freq_comparison}(a), we see that, for a lattice of rods, the frequencies of the single- and bi-domain lattices match, in accordance with the band diagram predictions (minor deviations are attributable to the small domain size, which departs from the infinite lattice conditions implicitly invoked in performing Bloch analysis). In contrast, for the lattice of beams in Fig.~\ref{Nat_freq_comparison}(b), the curves present an appreciable mismatch, which is especially conspicuous above $\approx$ 1.25 (range denoted by the yellow band), where the branches of the band diagrams for $\bar{\theta} = 110^o$ and $\bar{\theta} = 70^o$ start to deviate significantly, as recalled in inset (ii). This said, it is interesting to note that the mismatch in natural frequencies appears to be relatively modest across the spectrum, suggesting that, for problems involving the vibrations of bi-domain configurations, the qualitative manifestation of functional duality remains overall fairly immune from the property dilution induced by the loss of hinge ideality. 
	It is nonetheless worth pointing out that the bi-domain lattice considered in this comparison blends the frequency response characteristics of two configurations. If we were to compare a $6 \times 6$ lattice with $\bar{\theta} = 110^o$ against a $6 \times 6$ lattice with $\bar{\theta} = 70^o$ (offering an ``apple-to-apple" comparison of the dual configurations), the differences would be more accentuated, capturing more closely the deviations documented in inset (ii). This comparison is provided for completeness in the SM. In summary, this result suggests that, for vibration control problems, we can exploit the opportunities of stitched bi-domain arrangements, expecting reasonable preservation of functional duality even while working with structural lattices. Given the relevance of this class of problems for engineering applications, this robustness far from ideal conditions suggests a wide applicability of the duality paradigm.
	
	In conclusion, this study has elucidated that, in the presence of ideal bonds and connections, dual lattices can be effectively treated as dynamically equivalent solids, provided that proper stitching and material modulation are enforced. This is not necessarily the case for structural lattices, where a dilution of functional duality is observed, with different degrees of manifestation across wave propagation and vibration problems. It is worth emphasizing that, to some extent, a dilution of the nominal mechanical properties is an expected signature of the non-idealities resulting from structural hinges that is broadly observed in the mechanics of lattices. A pertinent example is the weakening and migration to finite frequencies of the floppy edge modes observed in topological lattices with ligament-like hinges, as discussed in~\cite{Ma-et-al_Topological-Waves_PRL_2018} (even though, in that case, the property dilution is mitigated by the topological protection, resulting in remarkable robustness of the edge localization properties far from ideal hinge conditions). However, the peculiar manifestations of such dilution presented in this study, i.e., the loss of behavioral symmetry in the configuration space of twisted kagome lattices and its consequences for wave propagation, are uniquely germane to the duality problem discussed herein.} \\
	
	
	
	\noindent The author acknowledges the support of the National Science Foundation (NSF Grant EFRI-1741618). 

\bibliography{bib}


	
		\newpage

	\section{Supplemental Information}
	
\section{Directivity mismatch between dual configurations}
\begin{figure} [!htb]
	\centering
	\includegraphics[scale=0.65]{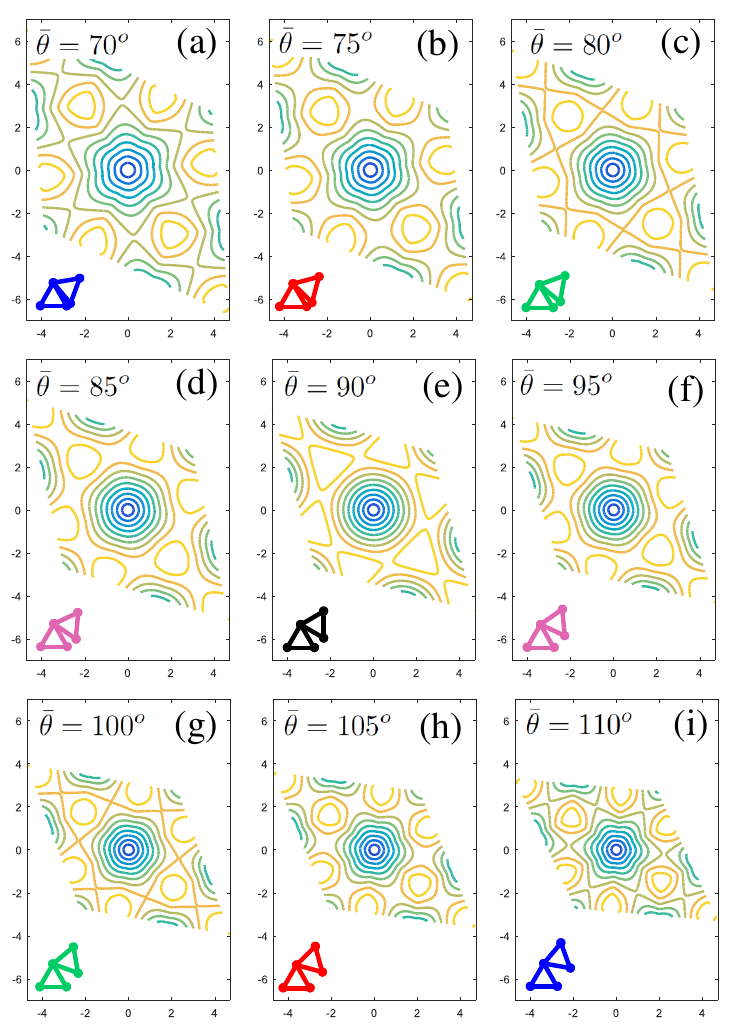}
	\caption{Iso-frequency contour lines for the I mode for lattice of rods connected by hinges for $\bar{\theta} \in [70^o \,\, 110^o]$.\\}
	\label{Isofrequency_Rods}
\end{figure}

In Fig.~\ref{Isofrequency_Rods}, we compare the iso-frequency contour lines for the I mode for lattices of rods connected by hinges, for $\bar{\theta} \in [70^o \,\, 110^o]$. If we compare dual configurations (e.g., (a) vs. (i)), we note that the magnitude of the wavevectors (and, consequently, wavelengths) and the directivity landscape are different, despite matching band diagrams.\\

\section{Effect of the slenderness ratio on the phonon duality for lattices of beams}
\begin{figure} [!htb]
	\centering
	\includegraphics[scale=0.52]{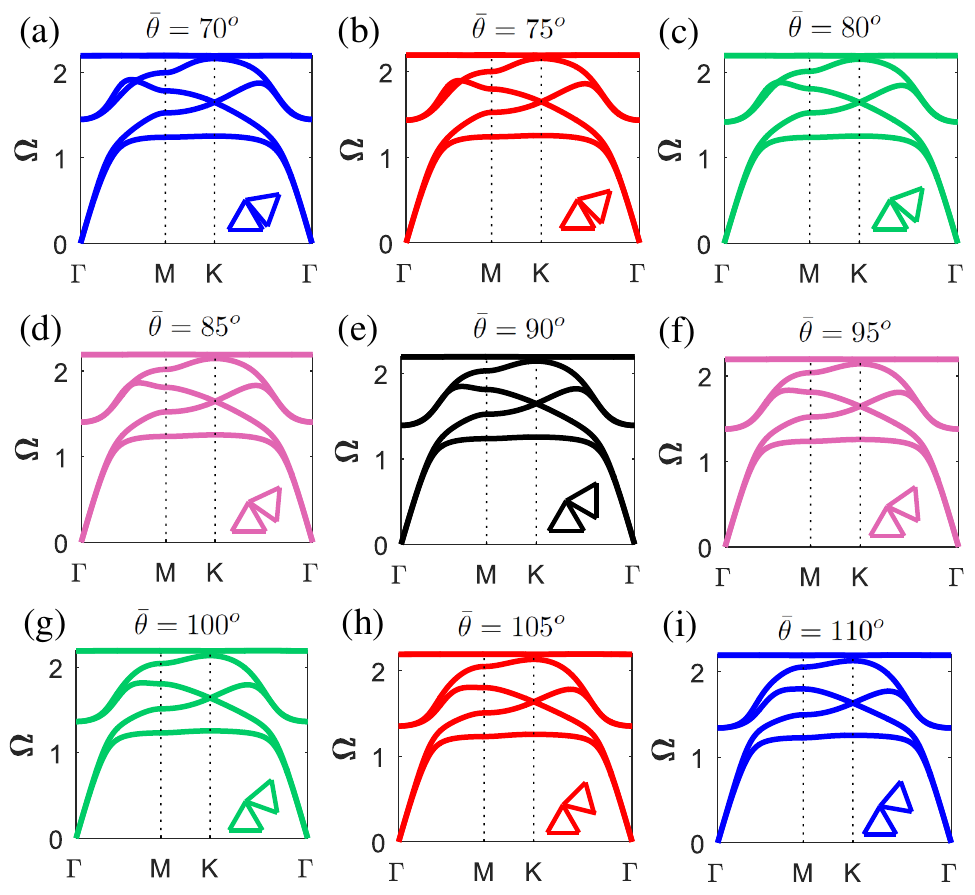}
	\caption{Band diagram for lattice of beams with slenderness ratio $\beta=1/15$, for $\bar{\theta} \in [70^o \,\, 110^o]$.}
	\label{BG_Beams_1over15}
\end{figure}

\begin{figure} [!htb]
	\centering
	\includegraphics[scale=0.52]{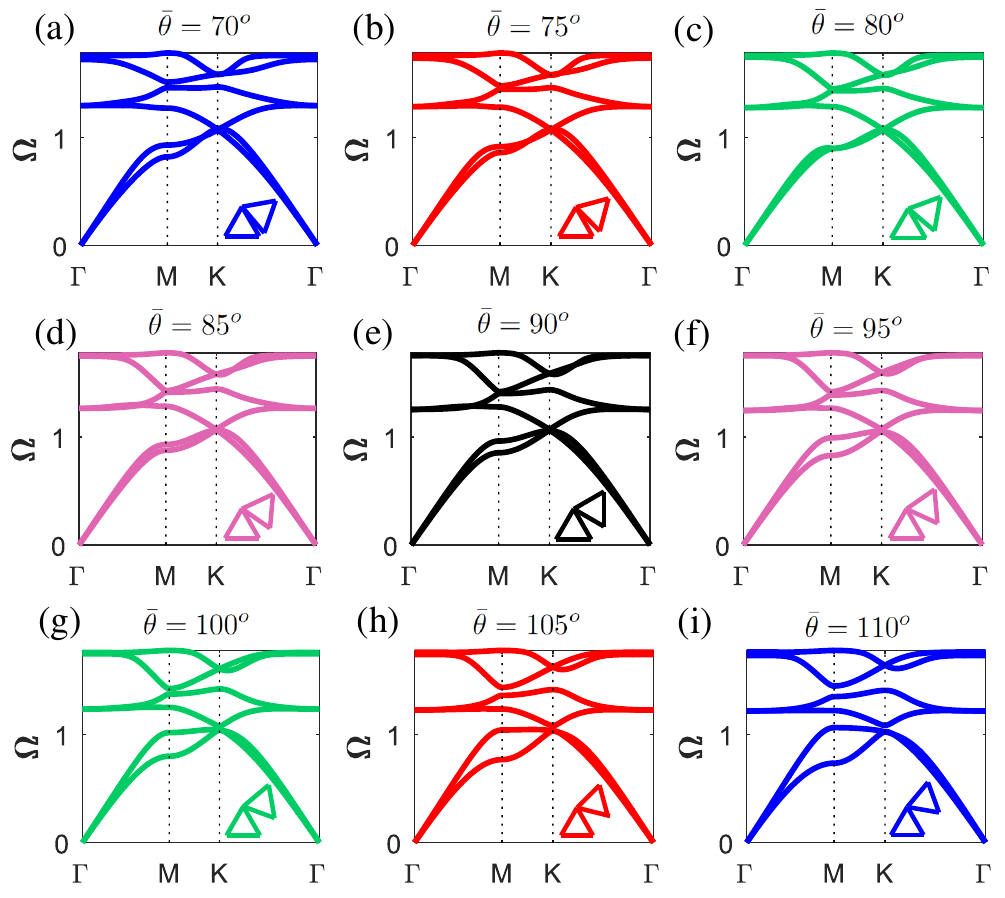}
	\caption{Band diagram for lattice of beams with slenderness ratio $\beta=1/5$, for $\bar{\theta} \in [70^o \,\, 110^o]$.}
	\label{BG_Beams_1over5}
\end{figure}

In Fig.s~\ref{BG_Beams_1over15} and~\ref{BG_Beams_1over5}, we plot the band diagrams for lattices of beams (with $\bar{\theta} \in [70^o \,\, 110^o]$) for slenderness ratio $\beta=1/15$ and $\beta=1/5$, respectively. We see that the differences between the phonon bands of dual configurations become more pronounced as $\beta$ increases, especially at low frequencies. The more we deviate from ideal lattice conditions (perfect hinges), the more the symmetry across the duality boundary is diluted.

\sg{\section{Note on the natural frequency modifications}}

\begin{figure} [!htb]
	\centering
	\includegraphics[scale=0.67]{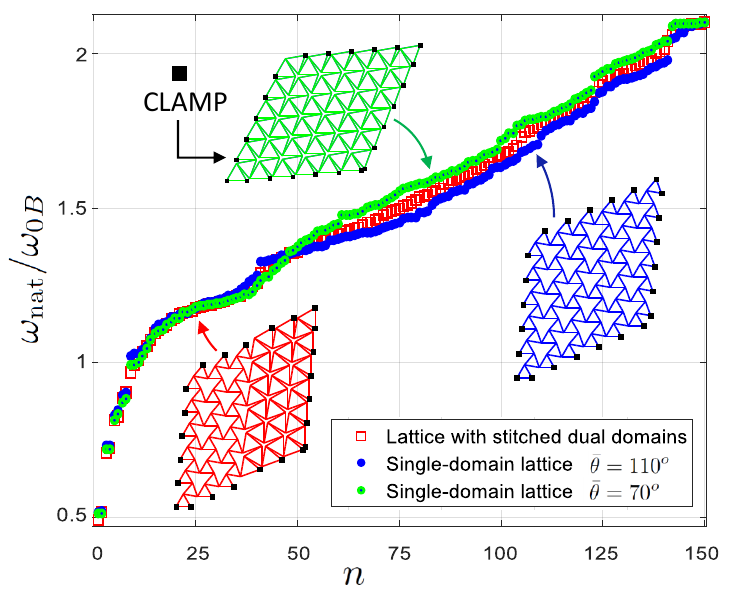}
	\caption{\sg{Comparison of natural frequencies between a bi-domain lattice with stitched dual configurations and two single-domain lattices with $\bar{\theta}=110^o$ and $\bar{\theta}=70^o$, respectively. The deviation between the two single-domain lattices is more pronounced than that between either one of them and the hybrid lattice, and better captures the dilution of functional duality associated with the transition from rods to beams.}}
	\label{Nat_freq_comparison_3-way}
\end{figure}

\sg{In Fig.~\ref{Nat_freq_comparison_3-way}, we compare the natural frequencies for three lattices of beams. The configurations color-coded in blue and red are the two discussed in the main article, i.e., the single-domain lattice with $\bar{\theta}=110^o$ and the lattice featuring two dual strips with $\bar{\theta}=110^o$ and $\bar{\theta}'=70^o$, respectively. The green set pertains to a single-domain lattice with $\bar{\theta}=70^o$. Since the red lattice is a hybrid structure that blends the characteristics of two configurations (and does not exactly correspond to the band diagram of either one), the spirit of this plot is to offer a more direct comparison of the two dual configurations, to match more closely the mismatch observed between their band diagrams. We observe that, indeed, the mismatch between blue and green is more pronounced than that between blue and red. While the comparison between blue and red, highlighted in the main article, is the most pertinent to our discussion on the potential of stitched domains to realize geometrically heterogeneous and dynamically homogeneous lattices, and reveals the robustness of such phenomenon even for lattices of beams, Fig.~\ref{Nat_freq_comparison_3-way} provides a more complete representation of the dilution of functional duality due to the transition from rods to beams.}

\section{Note on rod and beam elements}
\begin{figure} [b]
	\centering
	\includegraphics[scale=0.5]{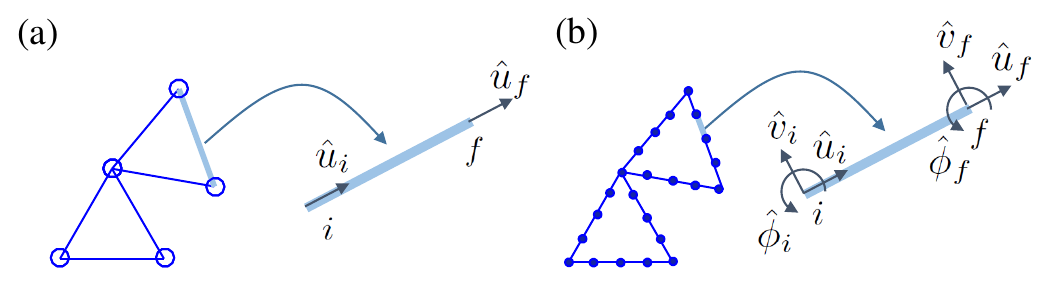}
	\caption{Rod and Timoshenko beam elements, showing their degrees of freedom in local coordinates.\\}
	\label{Rods_beams}
\end{figure}

A rod element is a two-node element with one degree of freedom per node, the axial displacements $\hat{u}_i$ and $\hat{u}_f$ expressed in local coordinates (see Fig.~\ref{Rods_beams}(a)). A rod element can only experience axial strains and stresses that are constant on the cross section. The stress-resultant forces are purely axial, such that a node at which several rods converge is subjected only to central forces. As no bending moments can be transmitted by rods, the hinges of a truss of rods have zero bending rigidity and therefore allow free relative rotation of the rods. 



For our beam meshes, we use Timoshenko beam elements. A Timoshenko beam element features three degrees of freedom per node: the axial nodal displacements $\hat{u}_i$ and $\hat{u}_f$ (as in the rod element), the lateral nodal displacements $\hat{v}_i$ and $\hat{v}_f$, also expressed in local coordinates, and the nodal rotational degrees of freedom $\hat{\phi}_i$ and $\hat{\phi}_f$ (see Fig.~\ref{Rods_beams}(b)) which represent cross-sectional rotations. The angular degree of freedom allows treating the tilt of the cross-section as an independent variable, reflecting the assumption (which becomes important for stubby beams), that the cross section may not remain perpendicular to the neutral axis of the beam during deformation. Beam elements can experience axial deformation as well as lateral deflection due to bending and shear. 
The resultant internal forces as well as moments are balanced at the lattice nodes, which behave as internal clamps that prevent relative rotation of the beams, implying that the angles formed by the beams meeting at a node are preserved locally (at the clamp) during deformation. 

Using a shear-deformable Timoshenko beam model is convenient because it guarantees a versatile platform against a variety of beam thicknesses (something that we exploit for the thickness sweep carried out in section II of this SM). However, the dichotomy of behavior between trusses and frames lies in the profound kinematic differences between rods and beams and is only marginally affected by the beam model that is adopted. Similar results could be achieved with more traditional Euler-Bernoulli beams. Incidentally, the Timoshenko model naturally yields Euler-Bernoulli results when the beam is sufficiently slender.
	
\end{document}